\documentclass[11pt]{article}
\usepackage{graphicx,xspace,amsmath,oupbib, changebar,url,wasysym}
\usepackage{booktabs}
\usepackage[italian,croatian,english]{babel}
\usepackage[latin1]{inputenc}
\usepackage{latexsym,color} 
\usepackage{epstopdf}
\usepackage{MnSymbol}
\usepackage{xcolor}
\usepackage{lineno}
\usepackage{tabularx}
\usepackage{epstopdf}

\newcommand{\ie}{\emph{i.e.}\xspace}

\newcommand{\be}{\begin{equation}}
\newcommand{\bel}[1]{\begin{equation}\label{#1}}
\newcommand{\ee}{\end{equation}}

\newcommand{\secref}[1]{\mbox{\S$\,$\ref{sec:#1}}}
\newcommand{\figref}[1]{\mbox{Figure~\ref{fig:#1}}}
\newcommand{\tabref}[1]{\mbox{Table~\ref{tab:#1}}}

\newcommand{\cd}{\,|\,}
\renewcommand{\H}{\mathcal{H}}

\newcommand{\LR}{\text{LR}}

\newcommand{\DNAmixtures}{{\tt DNAmixtures}}
\newcommand{\RHugin}{{\tt RHugin}}

\newcommand{\R}{{\tt R}}

\def\presuper#1#2%
{\mathop{}%
	\mathopen{\vphantom{#2}}^{#1}%
	\kern-\scriptspace%
	#2}

\begin{document}
	\title{{\sc Analysis of a  DNA
			mixture case involving  Romani reference populations}}
	\author{
		Francesco Dotto\thanks {Email: {\tt francesco.dotto@uniroma3.it}}
		\\
		Universit\`a Roma Tre, Italy.\\
		\and Julia Mortera\thanks {Email: {\tt julia.mortera@uniroma3.it}}
		\\
		Universit\`a Roma Tre, Italy.
		\and Laura Baldasarri\thanks{Email: {\tt laurabaldassarri72@gmail.com}}
		\\
		Universit\`a Cattolica del Sacro Cuore. 
		\and Vincenzo Pascali\thanks{Email: {\tt vincenzolorenzo.pascali@policlinicogemelli.it}}
		\\
		Universit\`a Cattolica del Sacro Cuore.
	}
\date{}
\maketitle
\begin{abstract}
	Here we present an Italian criminal case that shows how statistical methods can be used to extract information from a series of mixed DNA profiles. The case  involves several  different individuals and a set of different DNA traces. The case possibly involves persons of interest of a small, inbred population of Romani origin. First, a brief description of the case is provided. Secondly, we introduce some heuristic tools that can be used to evaluate the data and briefly outline the statistical model used for analysing DNA mixtures. Finally, we illustrate some of the findings on the case and discuss further directions of research.  The results show how the use of different population database  allele frequencies for analysing the DNA mixtures can lead to very different results, some seemingly inculpatory and some seemingly exculpatory. We also illustrate the results obtained from combining the evidence from different samples. 
\end{abstract}
\hspace{5mm}

\noindent {\small {\em Some key words:} Bayesian networks, combining evidence, DNA mixtures, forensic statistics, likelihood ratio, reference populations.}

\section{Introduction}

Here we present an Italian criminal case that shows how statistical methods can be used to extract information from a series of mixed DNA profiles. The case  involves several  different individuals and a set of different DNA traces. The case possibly involves persons of interest of a small, population of Romani origin. The  Romani or Romany, colloquially known as Gypsies or Roma, are an Indo-Aryan ethnic group, traditionally itinerant, living mostly in Europe and the Americas and originating from the Northern Indian subcontinent, \ie from Rajasthan, Haryana, and Punjab regions of modern-day India. We first give a brief overview of the presence of the Romani in Italy today. { Some historical background about this population is given in Appendix 1.} The case involved a certain number of persons of interest (PoI) and many DNA mixture traces.  A useful index is developed for the preliminary evaluation of which potential PoI is more likely to have contributed to a set of mixed DNA samples.  We base the analysis of the  DNA mixture on the model described in \textcite{cowell:etal:15}. This model takes fully into account the peak heights and the possible  artefacts, like stutter and dropout, that might occur in the DNA amplification process. The model is an extension of the gamma model developed in  \textcite{Gammamodel} and \textcite{cowell2007identification}, and used in  \textcite{cowell:etal:13}.    

The likelihood ratios computed using a fully continuous model led us to  subvert the inculpatory conclusions that were drawn by the public prosecutor's expert when using   semi-quantitative models. We also show that using appropriate population databases  is very important especially when   a genetically isolated population might be involved. 
In fact, the results show that  using   different population databases for the allele frequencies in analysing this case  can lead to very different results, some seemingly inculpatory and some seemingly exculpatory.  In \textit{People v. Prince case},\footnote{336 Cal.Rptr.3d 300 (Cal. Ct. App. 2005), rev. granted, 132 P.3d 210 (Cal. 2006), rev. dismissed,	142 P.2d 1184 (Cal. 2006).} a California Court of Appeal stated  that ``only the perpetrator's race be relevant to the crime; hence, it
is impermissible to introduce statistics about other races''. \textcite{kaye2008dna} rightly  critiques
this reasoning and presents a logical justification
for referring to a range of races and identifies  problems with the
one-race-only rule.

{ Here we also analyse the results when allowing for an ambient degree of relatedness among the Romani reference populations.}

\paragraph{The Romani in Italy}
 
There are roughly 130,000--170,000 people of  Romani and Sinti origin in Italy. They are about 0.23\% of the total population. This percentage is among the lowest in Europe. Roughly 50\% of the Romani and Sinti people in Italy are Italian citizens.
They are not a nomadic population: 85-90\% of the Romani and Sinti people in Europe have been sedentary for a long time. Only 2-3\%  in Italy are nomads. They are an extremely young  population: 60\%  are under 18 years of age, of those, 30\% are aged between 0 and 5 years, 47\% between 6 and 14, and 23\% are aged between 15 and 18.
Only 2-3\% are over 60.

\section{Outline of the Case}
\label{sec:case}

In a small village in North Italy, four men broke into a private courtyard trying to commit a theft. They were noticed  by two bystanders and fled.  The two bystanders alerted the local police station and, to escape from a patrol, the four rogues  stopped a car driver,  hijacked his car and then disappeared. The next day, the car was found, concealed in a country road, and a baseball cap was retrieved in the vehicle's seat. The cap did not belong to the car owner. The investigators concluded that the cap could be a link to identify one of the four offenders. The cap was brought to the local DNA laboratory, inspected under UV light and seven fabric samples were excised from its inner side. We will denote these samples B1, B2,$\cdots$,B7. The samples on the baseball cap were all DNA mixtures. 
Samples B1, B2, B4 and B5 were taken from the front of the cap, B7 from the crown and B3 and B6 from the underside of the cap peak.

Five individuals - two of them  of Romani ethnicity, together with the car owner  - were subsequently examined. We call these  persons of interest (PoI). A saliva swab was taken from all PoI to obtain their DNA profiles. These profiles were compared to the mixed  DNA evidence from  the baseball cap. 
 The analyst then concluded that,  on one hand, none of the contributors   matched  the consensus profile he  had previously given to the police. On the other hand,  he declared that he noticed a 
resemblance between the profiles on the cap and a profile of one of the six PoI.  We will henceforth, refer to this man as the  suspect $A$. The suspect was a middle aged man of Romani origin, {who had been previously condemned for other crimes.}
 A forensic scientist working for the police investigation, initially delivered  a written report on the case. The interpretation of the seven profiles were considered as  evidence  and a ``consensus''  genotype of an ``unknown'' individual was made.  The ``unknown'', was assumed to have had worn the cap while committing  the attempted robbery and had left his DNA on each of the seven mixed DNA samples, together with that of a couple of  other
  contributors unrelated to the robbery (the car owner's DNA was not  in any of the mixtures). However, the {consensus} genotype was incompatible with $A$'s genotype on 16 out of 21 loci.

 The statistical analysis {written in the report given to the court by the forensic scientist, involved}  seven samples and  was based on the use of four  distinct  software systems.  These were {ArmedXpert}\textsuperscript{\textregistered}, LRmixStudio, LabRetriever and DNAview, none of which use a fully continuous model for the peak height information {(unlike the study we give here)}, but use semi-continuous methods, which
are based on the allele information possibly in conjunction with probabilities of allelic dropout and dropin.  Seven separate likelihood ratios were reported for each analysis made with the different software systems. For sample B3, a likelihood ratio of $6.19 \times 10^6$ was given under a prosecution hypothesis that $A$ and $2$ unknowns contributed to the mixture. For sample B6, under  the same prosecution hypothesis a likelihood ratio of  $3.12 \times 10^{21}$  was reported.  {For both these likelihood ratios the alternative hypothesis was never mentioned. We argue that this is a strongly misleading way of reporting a likelihood ratio as both prosecution and defence hypotheses need to be clearly stated.}

In the report  the Caucasian\footnote{\texttt{https://strbase.nist.gov/NISTpop.htm}} and  an Italian  reference population of  allele frequencies  were used.  There was  no mention of the Romani ethnic group allele frequencies. { The suspect was from a small population of Romani origin, living in Piedmont,  North West Italy.  Romani   are an ethnic group  of about 10 million individuals of  predominantly West Eurasian ancestry scattered throughout Europe and  Asia \cite{moorjani2013reconstructing}.   Due to cultural and linguistic barriers, they seldom  intermingle  with Caucasians. Their allele frequencies are  therefore quite different from  Caucasians , and  Caucasian Italians. }   

As the aftermath of this court case --as might occur possibly in other judiciary proceedings-- the suspect was detained for several months awaiting  trial. He eventually agreed to plea a bargain, as he had already spent the time of a reduced sentence in jail. He was then sent for a further, short house detention. 

{ Here we show, what we believe should have been the way that this case should have been analysed. }

\section{Statistical Methods}
\label{sec:stat}
\subsection{Selection of PoI and samples}
\label{sec:poi}

The data available consists of seven different DNA samples, namely, B1, B2,$\cdots$, B7 and the DNA profiles of six people of interest, among whom $A$ the suspect. Here we give a preliminary heuristic evaluation on which potential contributor might have contributed to each of the 7 DNA samples. To do so, we built a   ``presence index" associated to each potential contributor within each DNA sample. 
{ This index  can be used in an investigative phase as a useful exploratory tool when one has many mixtures and potential contributors and no clear prosecution and defence hypotheses. This tool is useful to avoid having to compute all possible likelihood ratios for all combinations of contributors and mixtures for the analysis of  DNA mixtures using peak height information.}

Let $j=1,2,\dots,k$ denote the various DNA samples available, $m=1,\dots,M$ the different markers having $a=1,2\dots,A_m$  allelic types. For a marker $m$,   $n_{a}^m$ denotes the number of  alleles of type $a$ an individual possesses,  and   $O_m$ denotes the set of observed alleles in a mixture above a threshold $C$.  The  ``presence index" associated to each potential contributor to mixture $j$ is 

\begin{equation}
\label{eq:PI}
P =  \frac{1}{2M }\sum_{m=1}^M \sum_{a=1}^{A_m} I_{O_m}(n_{a}^m),
\end{equation}
 where
 $$I_{B}(x) = \begin{cases} 1 & \text{if} \; \; x \in B\\	0 & \text{otherwise} \end{cases}$$
  is the indicator function.

For mixture $j$, and potential contributor $i$, the index $P_{ij}$ 
\begin{equation}
P_{ij}=\begin{cases}
1 & \text{if all alleles possessed  by $i$ have a peak above $C$ } \\
0 & \text{if none of $i$'s alleles have a peak above $C$}.
\end{cases}
\end{equation}
 $P_{ij}$  measures the proportion of each potential contributor $i$'s alleles in mixture $j$. $P_{ij}$  ranges from $0$ to $1$.
 If none of  an individual $i$'s alleles are present in  mixture $j$  then  $P_{ij}=0$. {  Further details regarding the computation of index $P_{ij}$ are given in Appendix 2.}
 
 \begin{table}[ht]
{ 

	\caption{Presence index $P_{ij}$ for each  potential contributor  $i$ to each DNA mixture $j$.}	
		\label{tab:p.i}	
		\begin{center}
	\begin{tabular}{|c|rrrrrr|r|}
		
	& \multicolumn{6}{c|}{Persons of Interest} &  \\ 
			Samples  & $A$&$B$&$C$&$D$&$E$&$F$& Average\\
	 \hline
     B 1 & 0.50 & 0.62 & 0.50 & 0.35 & 0.38 & 0.42 & 0.46 \\ 
	 B 2 & 0.50 & 0.62 & 0.50 & 0.38 & 0.42 & 0.42 & 0.47 \\ 
     B 3 & 0.58 & 0.62 & 0.62 & 0.46 & 0.62 & 0.54 & 0.57 \\ 
     B 4 & 0.50 & 0.65 & 0.54 & 0.50 & 0.54 & 0.46 & 0.53 \\ 
	 B 5 & 0.50 & 0.62 & 0.50 & 0.35 & 0.38 & 0.42 & 0.46 \\ 
	 B 6 & 0.58 & 0.73 & 0.65 & 0.58 & 0.62 & 0.54 & 0.62 \\ 
	 B 7 & 0.50 & 0.62 & 0.58 & 0.38 & 0.42 & 0.50 & 0.50 \\ 
	\end{tabular}
\end{center}
}
\end{table}

The values of $P_{ij}$ are given  in  \tabref{p.i} and  indicate that the suspect $A$,  who was charged of the crime, is more likely to be a contributor to the mixtures B3 and B6, than to the other mixtures. Samples B3 and B6 are  those having the highest average presence index for suspect $A$. Individual $B$ has  the highest value of $P_{ij}$  among all the PoI in samples B3 and B6 and,  furthermore, $B$ has the highest value of $P_{ij}$ in these two samples.

 { The Presence index is  related to potential allelic drop out from the mixture. For example, consider the genotype of suspect $B$ in  sample B6. In this case, $P_{ij}$  is  0.73 ( the $6^{th}$ line of  \tabref{p.i}), which implies that $73\%$ of $B$'s alleles  are observed in   mixture B6. So, for a proposition that assumes  $B$ to be a contributor  to  mixture B6,   $27\%$ of his alleles would have to have dropped out during the amplification process. Thus,  the smaller $P_{ij}$, the smaller the likelihood that  individual $i$ is a contributor to the  mixture $j$.    \tabref{tab1}  shows the values, for the Italian reference population, of $P_{ij}$  and the corresponding likelihood ratio  $\LR$ (as computed in \secref{separate}) for  $H_p:  S_i\&U_1\&U_2$ versus  $H_d: U_1\&U_2\&U_3$ for mixture B6, where $S_i \in \{A,B,C,D,E,F\}$. Note that small values   of $P_{ij}$  correspond to  values of $\LR$ close to one, and large values   of $P_{ij}$  correspond to  large values of $\LR$.}

\begin{table}[ht]
	\begin{center}
		\caption{ Presence Index $P_{ij}$  for PoI,  $S_i \in\{A,B,C,D,E,F\}$, and corresponding likelihood ratio $\LR$ for $H_p:  S_i\&U_1\&U_2$  $H_d: U_1\&U_2\&U_3$  for the Italian population.}
		\label{tab:tab1}
		{
		\begin{tabular}{|c|rrrrrr|}
						& \multicolumn{6}{c|}{Persons of Interest}   \\ 
			& $A$&$B$&$C$&$D$&$E$&$F$\\	
			\hline
		    B 6 & 0.58 & 0.73 & 0.65 & 0.58 & 0.62 & 0.54  \\ 
			LR	& 1.04 & 44.91 & 1.87 & 1.44 & 9.91 & 1.03  \\ 
			
			\end{tabular}
		}
	\end{center}
\end{table}

\begin{figure}
	\centering
	 \resizebox{.75\textwidth}{!}{\includegraphics[angle=270]{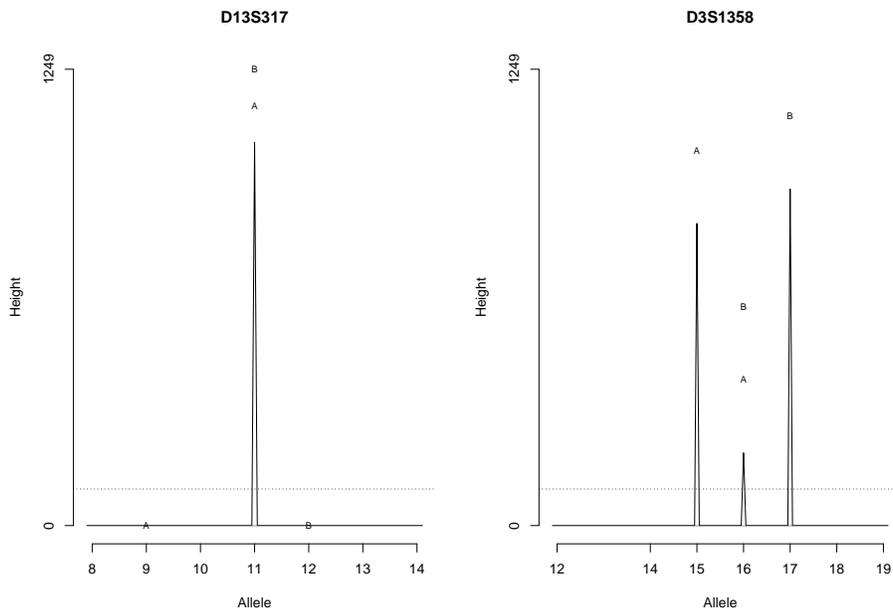}}
	\caption{A pictorial representation of the electropherogram, showing the peak heights and corresponding alleles,  for sample $B3$ at markers D13S317 and D3S1358. The genotypes of individuals  $A$ and $B$ at these markers are also given.}.
			\label{fig:epg}
\end{figure}

\figref{epg} shows a pictorial representation of the EPG for sample B3 at markers  D3S1358 and  D13S317. The labels $A$ and $B$ now denote the genotypes of the two suspects. Marker D13S317 in sample B3 has a single peak at allele $11$.  $A$ and $B$ have genotypes (9,11) and (11,12), respectively, so only their allele 11 is amplified in sample B3. So, if $A$  or $B$ were a contributor to B3, one of their alleles must not have been amplified, an artefact called a  dropout. For marker D3S1358, the EPG has peaks at alleles $15$, $16$, and $17$. $A$ has genotype (16,17) and $B$ has genotype (15,16), so all their alleles are present for this marker. However, under the hypothesis that both $A$ and $B$ are contributors, it is highly unlikely that the EPG would yield such a small peak at allele 16, as $A$ and $B$ would each contribute a proportion of DNA to the peak height at 16. Furthermore, if  $A$ and $B$ were present  their two alleles would be extremely  imbalanced.

\subsection{The weight of evidence}
\label{sec:woe}
Here we re-examine part of the evidence $E$ analysed with the semi-continuous methods used in the previous investigation.  This evidence consists of the peak heights and alleles in the EPGs of  samples B3 and B6, together with the genotypes of $A$ and $B$, \textit{i.e.} $E=\{A, B,\text{B3}, \text{B6}\}$. We consider how the evidence $E$ affects the comparison between the prosecution $H_p:  A\& U_1\&U_2 $  and  defence hypotheses $H_d:U_1\&U_2\&U_3 $
where, $H_p$ claims that $A$  and two unknown individuals, $U_1$ and $U_2$ contributed to the DNA mixture, whereas, $H_d$ states that 3 distinct  unknown individuals $U_1$, $U_2$ and $U_3$  contributed to the mixture. The DNA profiles of the known individuals are
considered  fixed, whereas the DNA profiles of the unknown contributors are considered mutually independent and sampled from a suitable
reference population. Similarly, a prosecution and defence hypothesis can be formulated for the other PoI, $B$, by substituting $B$ for $A$, in $H_p$. The strength of the evidence is reported as a likelihood ratio \cite{good1950probability,lindley1977problem}
\begin{equation}
\label{eq:likelihoodratio}
\LR= \frac{L(H_p)}{L(H_d)}=\frac{{P}(E|H_p)}{{P}(E|H_d)}
\end{equation} 
 or for large values of $\LR$  as the weight of evidence:  $\mbox{WoE} = \log_{10}\LR$  in the unit
 ban, so that 1 ban represents a factor 10 on
 the likelihood ratio \cite{good79}.

\subsection{Statistical model for DNA mixtures}
\label{sec:model}
We base the analysis of the  DNA mixture on the model described in \textcite{cowell:etal:15}. This model takes fully into account the peak heights and the possible  artefacts, like stutter and dropout, that might occur in the DNA amplification process. We give a brief summary of the main features of the model, for further details  we refer to \textcite{cowell:etal:15}. The model is an extension of the gamma model developed in  \textcite{Gammamodel} and \textcite{cowell2007identification}, and used in  \textcite{cowell:etal:13}. 

The model assumes that the variability at an allele is independent of the variability at other allelic positions
when the model parameters and genotypes are considered fully known. The model  takes into account artefacts: stutter, whereby a proportion of a peak belonging to allele $a$  appears as a peak at allele $a-1$; and dropout, when alleles are not observed because the peak height is below  a detection threshold $C$.

 Consider allele $a$, 
the variability of the  peak height $Z_a$  at $a$ can be expressed as the gamma
distribution 
\begin{equation}\label{eq:gammadist}
Z_{a} \sim \Gamma\left\{\frac{1}{\sigma^2}D_a(\phi, \xi, \mathbf{n}),\mu\sigma^2\right\},
\end{equation}
where $\phi_i$ denotes the \emph{proportion} of DNA originating
from individual $i$ prior to PCR amplification, $n_{ia}$ is the number of type $a$ alleles  for individual $i$,  $D_a(\phi, \xi, \mathbf{n})= (1-\xi)\sum_{i}\phi_in_{ia} + \xi\sum_{i}\phi_in_{i,a+1}$ are the effective allele counts after stutter. In a sample from a single donor, where no dropout or stutter has occurred,  $\mu$ is the mean peak height  and $\sigma$ is the coefficient of variation. For example, $\sigma = 0.58$ corresponds to the
standard deviation of the peak being 58\% of its mean $\mu$.   The
back-stutter parameter $\xi$ 
determines the mean proportion of stutter that may be observed in the
allelic position one repeat less. 
 Here $\xi$ is the ratio of the
stutter peak with respect to the parent plus the stutter peak, rather
than the more commonly used ratio between the stutter peak and the
parent peak \cite{tvedebrink2012allelic}.


The  evidence $E$ consists of the peak heights $\mathbf{z}$ as observed in the EPGs,  as well
as any potential genotypes of known individuals. For given genotypes of the contributors, expressed as allele counts $\mathbf{n}=(n_{ia}, i=1,\ldots  I; a =1,\ldots, A)$, given proportions $\phi$, and given values of the parameters $\psi=\left(\mu,\xi, \sigma \right)$, all observed
peak heights are independent and  for a given hypothesis $\H$,
the full likelihood is obtained by summing over all possible
combinations of genotypes $\mathbf{n}$  with probabilities $P(\mathbf{n}\cd \H)$ associated with $\H$: 
\begin{equation}
\label{eq:L}
L(\H) = \Pr(E\cd \H) = \sum_{\mathbf{n}} L(\mathbf{\psi} \cd \mathbf{z},\mathbf{n}) P(\mathbf{n}\cd \H),
\end{equation}
where \[L(\mathbf{\psi} \cd \mathbf{z}, \mathbf{n})=\prod_m\prod_a L_{ma}(z_{ma})\] and 
\begin{equation}\label{eq:dens}
L_{ma} (z_{ma})= \left\{ \begin{array}{cr} g\{z_{ma};\sigma^{-2} D_a(\phi, \xi,\mathbf{n}),\mu \sigma^2\}& \mbox{ if $z_{ma}\geq C$}\\G\{C; \sigma^{-2} D_a(\phi,\xi,\mathbf{n}), \mu\sigma^2\}&\mbox{otherwise,}\end{array}\right.
\end{equation}
with $g$ and $G$ denoting the gamma density and cumulative distribution function respectively. 

The number of terms in this sum is huge for a hypothesis which involves several unknown contributors to the mixture, but can be calculated efficiently by Bayesian network techniques that represent the genotypes using a Markovian structure, \ie\ the allele counts for each individual being modelled sequentially over the alleles. 
The maximum likelihood estimate (MLE) parameters are obtained using the \R package 
\DNAmixtures\ \cite{graversen:package:13}  which interfaces to the HUGIN API 
(Hugin Expert A/S, 2012) through the \R package \RHugin\ \cite{manual:RHugin}.

\subsection{Comparison of allele frequencies database}
\label{sec:plots}
 In both the prosecution and defence hypotheses that form the  likelihood ratio (\ref{eq:likelihoodratio}) one can have several unknown contributors from a  reference population. In order to compute the likelihood for each hypothesis in a DNA mixture (\ref{eq:L}) we need to compute the prior probabilities $P(\mathbf{n}\cd \H)$ associated with each hypothesis $\H$ by using a specific database of allele frequencies. This case concerns PoI from the Romani and Italian population, so we  used the following reference population (RP) allele frequency databases:  Macedonian Romani (M);  Portuguese Romani (PO); Eastern Slovakian Romani (ES)  \cite{havavs2007population,gusmao2010genetic,sotak2008genetic}, as well as the Italian Caucasian (IT). \tabref{number} shows the dimension of each of the reference population databases of allele frequencies.
 \begin{table}[ht!]
 	\centering
 	\caption{Size of the reference population databases of allele frequencies for the Romani and the Italian Caucasian population.}
 	\label{tab:number}
 	\begin{tabular}{r|ccc|c}
 		& \multicolumn{3}{c|}{Romani}& Italian \\
 		Reference Population  & Macedonian  & Portuguese  & Eastern Slovakian  & Caucasian \\
 		\hline
 		Dimension &        102 &        123 &        138 &        346 \\
 	\end{tabular}  
 \end{table}
 
An artificial admix reference population of the $3$ different Romani subpopulations was also made by forming a weighted average of the allele frequencies having weights equal to the dimensions given in \tabref{number}.
  
   \begin{figure}[ht!]
 	\caption{Allele frequency distribution for each population at 4 different markers: D18S51, D16S539, CSF1PO, FGA. $\leftrightline$ (black line) represents the Italian population, {\color{red}$\leftrightline $  } (red line) represents the Macedonian Romani population, {\color{green} $\leftrightline$  } (green line) represents the Eastern Slovakian Romani population, {$\leftrightline$ } (blue line) represents the Portuguese Romani population while {\color{cyan} $\leftrightline$ } represents the ``Mixed Romani population''. }
 	\label{fig:allele.freq}
 	\includegraphics[width=0.95\textwidth]{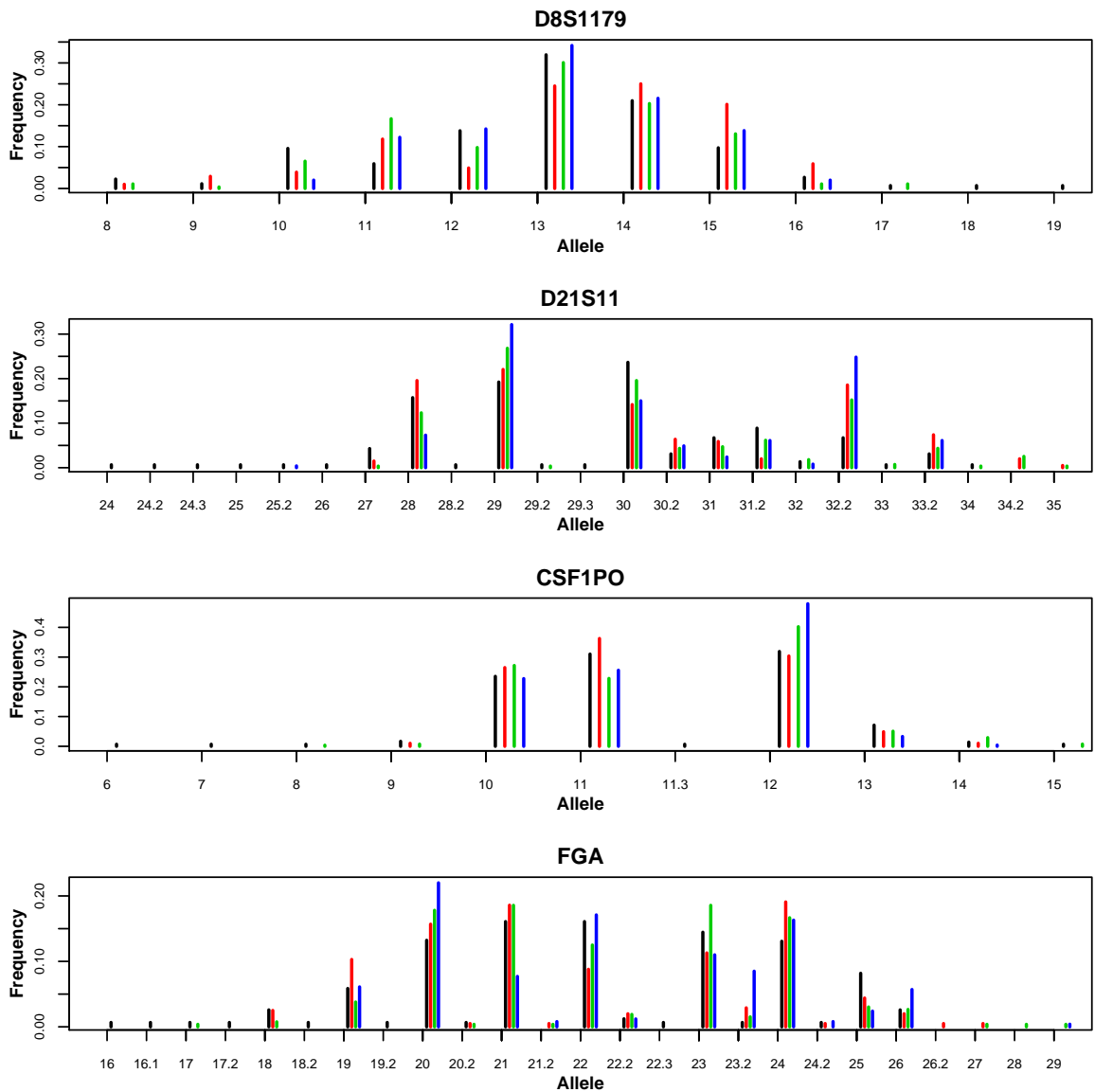}
 \end{figure}

  The allele frequency distributions for markers D8S1179, D21S11, CSF1PO and FGA are shown in Figure \ref{fig:allele.freq}. Some of the Romani allele frequencies are  different from the Caucasian allele frequencies. Take for example, markers D21S11 and FGA, where the Romani populations tend to have large allele values not present among the Caucasian database, whereas the Caucasian have small allele values not observed among the Romani. However, the Caucasian database  is much bigger, being based on over 346 individuals compared to the hundred odd individuals in the Romani databases. The  smaller alleles present in the Caucasian database being rare might not have been observed in the  Romani databases.

\section{Results}
\label{sec:results}
The evidence $E$ consists of the peak heights and alleles in two EPGs from  samples B3 and B6, together with the genotypes of the suspect $A$ and another individual $B$, \textit{i.e.} $E=\{A,B,\text{B3}, \text{B6}\}$. We consider how this evidence $E$ affects the comparison between the prosecution $H_p:  A\& U_1\&U_2 $  and  defence hypotheses $H_d:U_1\&U_2\&U_3 $
where, $H_p$ claims that $A$  and two unknown individuals, $U_1$ and $U_2$ contributed to the DNA mixture, whereas, $H_d$ states that 3 distinct  unknown individuals $U_1$, $U_2$ and $U_3$  contributed to the mixture. Similarly, a prosecution and defence hypothesis can be formulated for the other PoI, $B$, by substituting $B$ for $A$, in $H_p$. 
Here we analyse the evidence in the two samples separately and in combination. In all the analyses the standard threshold value $C=50$ is used. The additional unknown contributor in each hypothesis can account for any dropin that might occur in the amplification process.

\subsection{Separate analysis of samples B3 and B6.}
\label{sec:separate}
This case concerns PoI from the Romani and Italian population, so we  used the following reference population (RP) allele frequency databases:  Macedonian Romani (M);  Portuguese Romani (PO); Eastern Slovakian Romani (ES) as well as the Italian Caucasian (IT). 
   
 \tabref{A} shows the parameter estimates and  the likelihood ratios for comparing the hypotheses  $H_p: A\&U_1\&U_2$ \mbox{\textit{vs.}} $H_d: U_1\&U_2\&U_3$ for  the DNA mixtures in samples $B3$ and $B6$ using the 4 different reference populations.

\begin{table}[ht!]
	\centering
	\caption{Estimated parameters for DNA mixture samples B3 and B6 under the prosecution hypothesis $H_p:  A\&U_1\&U_2$  and a defence hypothesis $H_d: U_1\&U_2\&U_3$.}
\label{tab:A}
	\small
	\begin{tabular}{|l|rrrrrr|rrrrrr|}
			\hline
		&   \multicolumn{12}{c|}{Sample B3} \\ 
		&   \multicolumn{6}{c|}{$H_p:  A\&U_1\&U_2$}    & \multicolumn{6}{c|}{$H_d: U_1\&U_2\&U_3$}  \\
		\hline
	    RP & $\mu$ & $\sigma$ & $\xi$ & $\phi_{U_1}$ &$\phi_{U_2}$ & $\phi_{A}$  & $\mu$ & $\sigma$ & $\xi$ & $\phi_{U_1}$ & $\phi_{U_2}$ & $\phi_{U_3}$ \\ 
	    \hline
		IT & 523 & 0.88 & 0.00 & 0.55 & 0.30 & 0.16 & 517 & 0.84 & 0.00 & 0.68 & 0.32 & 0.00 \\ 
		MA & 521 & 0.86 & 0.00 & 0.45 & 0.45 & 0.10 & 518 & 0.84 & 0.00 & 0.50 & 0.50 & 0.00 \\ 
		PO & 520 & 0.86 & 0.00 & 0.45 & 0.45 & 0.10 & 516 & 0.84 & 0.00 & 0.50 & 0.50 & 0.00 \\ 
		ES & 536 & 0.90 & 0.00 & 0.42 & 0.42 & 0.15 & 531 & 0.86 & 0.00 & 0.50 & 0.50 & 0.00\\			\hline
		&   \multicolumn{12}{c|}{Sample B6} \\ \hline 
	RP & $\mu$ & $\sigma$ & $\xi$ & $\phi_{U_1}$ &$\phi_{U_2}$ & $\phi_{A}$  & $\mu$ & $\sigma$ & $\xi$ & $\phi_{U_1}$ & $\phi_{U_2}$ & $\phi_{U_3}$ \\ 
		\hline
    IT & 528 & 0.65 & 0.10 & 0.68 & 0.28 & 0.04 & 525 & 0.65 & 0.10 & 0.68 & 0.32 & 0.00 \\ 
	MA & 526 & 0.68 & 0.13 & 0.68 & 0.30 & 0.02 & 529 & 0.68 & 0.12 & 0.69 & 0.16 & 0.16 \\ 
	PO & 527 & 0.68 & 0.13 & 0.65 & 0.28 & 0.06 & 527 & 0.68 & 0.11 & 0.65 & 0.18 & 0.18 \\ 
	ES & 531 & 0.70 & 0.18 & 0.69 & 0.17 & 0.14 & 526 & 0.69 & 0.14 & 0.69 & 0.31 & 0.01 \\ 
			\hline
	\end{tabular}
\end{table}

\begin{table}[ht!]
\centering
\caption{Likelihood ratios for   $H_p:  A\&U_1\&U_2$ \mbox{\textit{vs.}} $H_d: U_1\&U_2\&U_3$ for the separate analysis of the DNA samples B3 and B6 in the four different reference populations.}
	\label{tab:ALR}	
		\begin{tabular}{r|rrrr|rrrr}
					& \multicolumn{4}{c|}{B3} &       \multicolumn{4}{c}{B6}\\	\hline
Reference Population &         IT &         MA &         PO &         ES &         IT &         MA &         PO &         ES \\
			\LR &    1.80& 1.22 & 1.38 & 1.75& 1.04& 0.99 & 1.00 & 1.83\\
					\end{tabular}  
\end{table}

\tabref{A} and \tabref{B} show that the  parameter estimates for $\mu$ and $\sigma$ { are similar in} in the different reference populations and the different hypotheses. This is expected as these parameters do not depend on the reference population used but only on the peak heights in the EPGs. Furthermore, the stutter parameter $\xi$ is almost null, indicating that potential stutter had been filtered out of the data. This can be dangerous as a true peak might be confused with a stutter peak. 
 
 The proportion of DNA contributed by the main suspect $A$ in \tabref{A} is small both in mixture B3, $\phi_{A} \in [0.1, 0.16]$, and even smaller in mixture B6, $\phi_{A} \in  [0.02, 0.14]$. Under the prosecution proposition $A$ is always estimated to be the minor contributor to the DNA mixtures. 
 
  The likelihood ratio \LR\ in \tabref{ALR} for comparing $H_p: A\&U_1\&U_2$ \mbox{\textit{vs.}} $H_d: U_1\&U_2\&U_3$ for the separate analysis of the samples yields a maximum value  $\LR =1.75$ for B3 and $\LR = 1.83$ for B6  when using the Eastern Slovakian reference population. This result is far from being incriminating for the suspect $A$, contrary to the investigative analysis that led to the suspect being detained (see \secref{case}).

 \begin{table}[ht!]
 	\centering
 		\caption{Estimated parameters for DNA mixture samples B3 and B6 under the prosecution hypothesis $H_p:  B\&U_1\&U_2$  and a defence hypothesis $H_d: U_1\&U_2\&U_3$.}
 	\label{tab:B}
 	\small
  	\begin{tabular}{|l|rrrrrr|rrrrrr|}
 		\hline
 		&   \multicolumn{12}{c|}{Sample B3} \\ 
 		&   \multicolumn{6}{c|}{$H_p:  B\&U_1\&U_2$}    & \multicolumn{6}{c|}{$H_d: U_1\&U_2\&U_3$}  \\
 		\hline
 		RP & $\mu$ & $\sigma$ & $\xi$ & $\phi_{U_1}$ &$\phi_{U_2}$ & $\phi_{B}$  & $\mu$ & $\sigma$ & $\xi$ & $\phi_{U_1}$ & $\phi_{U_2}$ & $\phi_{U_3}$ \\ 
 		\hline
 	IT & 524 & 0.88 & 0.00 & 0.57 & 0.20 & 0.24 & 517 & 0.84 & 0.00 & 0.68 & 0.32 & 0.00 \\ 
 	MA & 524 & 0.87 & 0.00 & 0.48 & 0.31 & 0.21 & 518 & 0.84 & 0.00 & 0.50 & 0.50 & 0.00 \\ 
 	PO & 522 & 0.86 & 0.00 & 0.49 & 0.35 & 0.16 & 516 & 0.84 & 0.00 & 0.50 & 0.50 & 0.00 \\ 
 	ES & 523 & 0.89 & 0.00 & 0.53 & 0.25 & 0.22 & 517 & 0.85 & 0.00 & 0.57 & 0.43 & 0.00 \\ 
 		\hline
 		&   \multicolumn{12}{c|}{Sample B6} \\ 
 		\hline
 		RP & $\mu$ & $\sigma$ & $\xi$ & $\phi_{U_1}$ &$\phi_{U_2}$ & $\phi_{B}$  & $\mu$ & $\sigma$ & $\xi$ & $\phi_{U_1}$ & $\phi_{U_2}$ & $\phi_{U_3}$ \\ 
 		\hline
 		IT & 532 & 0.69 & 0.10 & 0.58 & 0.14 & 0.29 & 524 & 0.65 & 0.08 & 0.67 & 0.33 & 0.00 \\ 
 		MA & 531 & 0.73 & 0.12 & 0.55 & 0.18 & 0.27 & 529 & 0.68 & 0.12 & 0.69 & 0.16 & 0.16 \\ 
 		PO & 529 & 0.72 & 0.11 & 0.52 & 0.24 & 0.24 & 527 & 0.68 & 0.11 & 0.65 & 0.18 & 0.18 \\ 
 		ES & 532 & 0.75 & 0.15 & 0.55 & 0.15 & 0.30 & 526 & 0.69 & 0.14 & 0.69 & 0.31 & 0.01 \\ 
 		\hline
 	\end{tabular}
 \end{table}

 \begin{table}[ht!]
 	\centering
 	 	\caption{Likelihood ratios for   $H_p:  B\&U_1\&U_2$ \mbox{\textit{vs.}} $H_d: U_1\&U_2\&U_3$ for the separate analysis of the DNA samples B3 and B6 in the four different reference populations.}
 \label{tab:BLR}
 	\begin{tabular}{r|rrrr|rrrr}
 		& \multicolumn{4}{c|}{B3} &       \multicolumn{4}{c}{B6}\\	\hline
 		Reference Population &         IT &         MA &         PO &         ES &         IT &         MA &         PO &         ES \\
 		\LR&	24.16 &      9.38  &      2.91  &     19.94  &     44.91 &       23.58&       7.50&      54.33 \\
 		\hline
 	\end{tabular}  
 \end{table}

  However, in \tabref{B} the proportion of DNA contributed by the PoI $B$, is  larger than that for the suspect $A$, $\phi_{B}\in [0.16, 0.24]$ for mixture B3 and  $\phi_{B}\in [0.24, 0.30]$ for mixture B6. Furthermore, the $\LR$s in \tabref{BLR} for comparing $H_p: B\&U_1\&U_2$ \mbox{\textit{vs.}} $H_d: U_1\&U_2\&U_3$ are much larger than that for suspect $A$. In sample B3 the maximum is $\LR=24.16$ for the Italian Caucasian RP, whereas in B6 the maximum is $\LR=54.33$ for the Eastern Slovakian Romani RP, indicating that $B$ might be a minor contributor to both mixtures. For the Portuguese Romani RP,  the $\LR=2.91$ is small for sample B3 and is $\LR=7.5$ for sample B6.   This illustrates how the weight of evidence can vary when using different reference populations, some potentially leading to exonerating and some to condemning a suspect. 
This case shows that the Romani population might still be discriminated against in modern day Italy.

\subsection{Combining the evidence from samples B3 and B6}
\label{sec:joint}

There are many reasons for combining evidence, one important reason being
that it strengthens the information about the profiles of any shared
contributors. 
Combining the information in multiple profiles requires a slightly
more complex analysis than that of single DNA mixture profiles, since it is now necessary to make assumptions about which -- if any -- contributors
may be in common \cite{graversen2019yara,pascali2012joint}. When combining replicates it is natural to make an
assumption that contributors are the same, however when combining
profiles from different samples one needs to carefully consider
whether there is perhaps only a partial overlap. However, once a
hypothesis describing the contributors is formulated, the mathematical
details in extending the model from one to multiple crime
scene profiles is completely straightforward.
One can assume in fact that, conditionally on the DNA profiles of the entire pool
of contributors (and the model parameters), the peak heights in one
EPG are independent of the peak heights in the other EPGs.  

\tabref{Aunc} gives the parameter estimates for the combined analysis  of samples B3 and B6 under the prosecution proposition
$H_p:  A\&U_1\&U_2$  and a defence proposition $H_d: U_1\&U_2\&U_3$, where we assume that the unknown contributors may be in common.

 \begin{table}[ht!]
\centering
	\caption{Estimated parameters for the joint analysis of  DNA mixture samples B3 and B6 under the prosecution hypothesis $H_p: A\&U_1\&U_2$  and a defence hypothesis $H_d: U_1\&U_2\&U_3$.}
	\label{tab:Aunc}
	\small
		\begin{tabular}{|l|rrrrrr|rrrrrr|}
		\hline
		&   \multicolumn{12}{c|}{Sample B3} \\ 
		&   \multicolumn{6}{c|}{$H_p:  A\&U_1\&U_2$}    & \multicolumn{6}{c|}{$H_d: U_1\&U_2\&U_3$}  \\
		\hline
		RP & $\mu$ & $\sigma$ & $\xi$ & $\phi_{U_1}$ &$\phi_{U_2}$ & $\phi_{A}$  & $\mu$ & $\sigma$ & $\xi$ & $\phi_{U_1}$ & $\phi_{U_2}$ & $\phi_{U_3}$ \\ 
		\hline
		IT & 530 & 0.74 & 0.00 & 0.82 & 0.11 & 0.07 & 531 & 0.75 & 0.00 & 0.82 & 0.09 & 0.09 \\ 
		MA & 530 & 0.74 & 0.00 & 0.81 & 0.11 & 0.08 & 531 & 0.74 & 0.00 & 0.81 & 0.09 & 0.09 \\ 
		PO & 529 & 0.74 & 0.00 & 0.81 & 0.10 & 0.09 & 530 & 0.75 & 0.00 & 0.80 & 0.10 & 0.10 \\ 
		ES & 529 & 0.74 & 0.00 & 0.82 & 0.08 & 0.10 & 530 & 0.75 & 0.00 & 0.82 & 0.09 & 0.09 \\
		&   \multicolumn{12}{c|}{Sample B6} \\ 
	\hline 
		RP & $\mu$ & $\sigma$ & $\xi$ & $\phi_{U_1}$ &$\phi_{U_2}$ & $\phi_{A}$&  $\mu$ & $\sigma$ & $\xi$ & $\phi_{U_1}$ & $\phi_{U_2}$ & $\phi_{U_3}$  \\ 
		\hline
		IT & 543 & 0.55 & 0.01 & 0.73 & 0.27 & 0.00 & 548 & 0.56 & 0.00 & 0.69 & 0.16 & 0.16 \\ 
		MA & 554 & 0.57 & 0.10 & 0.76 & 0.17 & 0.07 & 551 & 0.57 & 0.05 & 0.72 & 0.14 & 0.14 \\ 
		PO & 555 & 0.57 & 0.12 & 0.76 & 0.14 & 0.10 & 550 & 0.57 & 0.06 & 0.71 & 0.15 & 0.15 \\ 
		ES & 555 & 0.59 & 0.18 & 0.82 & 0.00 & 0.18 & 552 & 0.58 & 0.08 & 0.74 & 0.13 & 0.13 \\ 
		\hline
	\end{tabular}
\end{table}

Comparing \tabref{Aunc}  to \tabref{A} we see that the variability $\sigma$ decreases and the proportions given by each contributor to the mixture are more clear cut. This shows that combining the evidence can lead to a better inference about the mixture. Similar results can be seen for the  hypotheses involving $B$,  when comparing \tabref{Bunc} for the joint analysis  to \tabref{B} for the separate analysis of B3 and B6.
\begin{table}[ht!]
	\centering
	\caption{Estimated parameters for the joint analysis of  DNA mixture samples B3 and B6 under the prosecution hypothesis $H_p: B\&U_1\&U_2$  and a defence hypothesis $H_d: U_1\&U_2\&U_3$.}
	\label{tab:Bunc}
	\small
	\begin{tabular}{|l|rrrrrr|rrrrrr|}
		\hline
		&   \multicolumn{12}{c|}{Sample B3} \\ 
		&   \multicolumn{6}{c|}{$H_p:  B\&U_1\&U_2$}    & \multicolumn{6}{c|}{$H_d: U_1\&U_2\&U_3$}  \\
		\hline
		RP & $\mu$ & $\sigma$ & $\xi$ & $\phi_{U_1}$ &$\phi_{U_2}$ & $\phi_{B}$  & $\mu$ & $\sigma$ & $\xi$ & $\phi_{U_1}$ & $\phi_{U_2}$ & $\phi_{U_3}$ \\ 
		\hline
		IT & 532 & 0.76 & 0.00 & 0.77 & 0.17 & 0.06 & 531 & 0.75 & 0.00 & 0.82 & 0.09 & 0.09 \\ 
		MA & 529 & 0.75 & 0.00 & 0.81 & 0.19 & 0.00 & 531 & 0.74 & 0.00 & 0.81 & 0.09 & 0.09 \\ 
		PO & 528 & 0.76 & 0.00 & 0.81 & 0.19 & 0.00 & 530 & 0.75 & 0.00 & 0.80 & 0.10 & 0.10 \\ 
		ES & 531 & 0.76 & 0.00 & 0.78 & 0.16 & 0.06 & 530 & 0.75 & 0.00 & 0.82 & 0.09 & 0.09 \\
		\hline
		&  & &  &  &  \multicolumn{5}{c}{Sample B6} &  &   & \\ 
		\hline  
		RP & $\mu$ & $\sigma$ & $\xi$ & $\phi_{U_1}$ &$\phi_{U_2}$ & $\phi_{B}$&  $\mu$ & $\sigma$ & $\xi$ & $\phi_{U_1}$ & $\phi_{U_2}$ & $\phi_{U_3}$  \\ 
		\hline
		IT & 548 & 0.55 & 0.00 & 0.64 & 0.25 & 0.11 & 548 & 0.56 & 0.00 & 0.69 & 0.16 & 0.16 \\ 
		MA & 549 & 0.56 & 0.04 & 0.69 & 0.24 & 0.07 & 551 & 0.57 & 0.05 & 0.72 & 0.14 & 0.14 \\ 
		PO & 550 & 0.56 & 0.06 & 0.69 & 0.24 & 0.08 & 550 & 0.57 & 0.06 & 0.71 & 0.15 & 0.15 \\ 
		ES & 554 & 0.58 & 0.09 & 0.69 & 0.17 & 0.14 & 552 & 0.58 & 0.08 & 0.74 & 0.13 & 0.13 \\ 
		\hline
	\end{tabular}
\end{table}

\begin{table}[ht!]
	\centering
	\caption{Likelihood ratios \LR\ for   $H_p:  A\&U_1\&U_2$ \mbox{\textit{vs.}} $H_d: U_1\&U_2\&U_3$ for the combined analysis of the DNA samples B3 and B6 in the four different reference populations when the contributors are in common and when they are distinct.}
		\label{tab:ALRunc}
		\begin{tabular}{r|rrrr}
			Reference Population &         IT &         MA &         PO &         ES \\
		\hline
Common unknown contributors	 & 1.20 & 0.77 & 2.83 & 2.5 \\ 
Distinct unknown contributors	 & 1.87 & 1.20 & 1.38 & 3.2 \\
	\end{tabular}
\end{table}

The first row of \tabref{ALRunc} gives  the \LR\  for   $H_p:  A\&U_1\&U_2$ \mbox{\textit{vs.}} $H_d: U_1\&U_2\&U_3$ for the combined analysis of the DNA samples B3 and B6 in the four different reference populations when the contributors may be in common in the two mixtures. The second row of \tabref{ALRunc} gives the combined analysis when we consider the contributors as distinct. The latter \LR\ is obtained by independence simply as the product of the $\LR$s  for B3 and B6 given in \tabref{ALR}. 

Similarly, \tabref{BLRunc} gives  the \LR\  for   $H_p:  B\&U_1\&U_2$ \mbox{\textit{vs.}} $H_d: U_1\&U_2\&U_3$  for the combined analysis of the DNA samples B3 and B6 when the contributors are considered in common and distinct. Note that in this case, considering the contributors as distinct yields a much larger \LR\ than when the contributors are in common. For example, when considering the Eastern Slovakian reference population under common contributors the $\LR=0.46$, whereas, for distinct contributors to B3 and B6 the $\LR= 1083$, which is over 2300 times  that when the contributors are in common.
This illustrates that care should be taken in formulating the hypotheses on which contributors are in common and which are distinct in the joint analysis of two or more mixtures.

\begin{table}[ht!]
	\centering
	\caption{Likelihood ratios \LR\ for   $H_p:  B\&U_1\&U_2$ \mbox{\textit{vs.}} $H_d: U_1\&U_2\&U_3$ for the combined analysis of the DNA samples B3 and B6 in the four different reference populations when the contributors are in common and when they are distinct.}
	\label{tab:BLRunc}
	\begin{tabular}{r|rrrr}
		Reference Population &         IT &         MA &         PO &         ES \\
		\hline
		Common unknown contributors	  &0.81 & 0.58 & 0.70 & 0.46 \\  
		Distinct unknown contributors &  1085 &221 & 21 &1083 \\	
	\end{tabular}
\end{table}

{\section{Population relatedness}
The standard approach to allowing for an ambient
degree of relatedness in a population is by means of the coancestry coefficient  $\theta$, which corresponds to  Wright's measure of interpopulation variation $F_{ST}$ \cite{wright1940breeding}.
Here we give  results when assuming that there is an ambient degree of relatedness in the Romani population, by applying the methodology for  DNA mixtures illustrated by Peter Green  in his comment to \textcite{cowell:etal:15}, where he shows that in this scenario, when genotypes are represented by allele count arrays,  the individual allele counts have a Beta-Binomial conditional distributions. This model was also used by \textcite{tvedebrink2010} for modelling kinship. } 

{ The results are shown in the  Table \ref{tab:UAFLR}. Note that  population relatedness  has only a very slight impact on the results.Similar results (not shown here) were found for the hypotheses concerning the PoI $B$. As noted by Peter Green for other examples, here too this
uncertainty can either increase (as in mixture B3) or decrease (as in mixture B6) the likelihood ratio $\LR$. This is in contrast with  earlier examples using mixture models only involving discrete allele presence
\cite{green:mortera:09} or for the model in  Cowell et al. (2007). In both these cases,  this uncertainty
always reduced the weight of evidence.}

\begin{table}[ht!]
	\centering
	\caption{Likelihood ratios for   $H_p:  A\&U_1\&U_2$ \mbox{\textit{vs.}} $H_d: U_1\&U_2\&U_3$ for the separate analysis of the DNA samples B3 and B6 in the four different reference populations.}
	\label{tab:UAFLR}	
{ 	\begin{tabular}{r|rrr|rrrr}
		& \multicolumn{3}{c|}{B3} &       \multicolumn{3}{c}{B6}\\	
		Reference Population &                 MA &         PO &         ES &                  MA &         PO &         ES \\
		\hline
		LR - No relatedness     &  1.22 & 1.38 & 1.75&  0.99 & 1.00 & 1.83\\
		LR - $\theta=0.05$      & 1.19 & 1.35 & 1.74&  0.96 & 1.01& 1.69 \\ 
			LR - $\theta=0.02$  & 1.12 & 1.28 &1.68 &   1.06 & 1.11 &  1.77  \\ 
			LR - $\theta=0.01$  &  1.11 & 1.26 & 1.68&  1.10  & 1.15 &  1.81  \\ 
		
	\end{tabular}  }
\end{table}

%
%
%

\section{Conclusions}
\label{sec:conc}

\paragraph{Admixed Population}

Using an admix reference population   made by forming a weighted average of the allele frequencies of the $3$ different Romani subpopulations, having weights equal to the subpopulation dimensions in \tabref{number}, we obtain the likelihood ratios shown in \tabref{admix2}.

\begin{table}[ht]
		\centering
	\caption{Likelihood ratios for prosecution hypothesis with suspect $A$ and PoI $B$, $H_p:  A\&U_1\&U_2$ \mbox{\textit{vs.}} $H_d: U_1\&U_2\&U_3$  and  $H_p: B\&U_1\&U_2$ \mbox{\textit{vs.}} $H_d: U_1\&U_2\&U_3$, for the admixed population for both the separate and combined analysis of mixtures B3 and B6.}
	\label{tab:admix2}
	\begin{tabular}{r|rr|rr|rr}
		
		Sample & \multicolumn{2}{c|}{B3} &  \multicolumn{2}{c|}{B6} &  \multicolumn{2}{c}{B3 \& B6 combined} \\ 
						Suspect	& $A$ & $B$ & $A$ & $B$ & $A$ & $B$ \\ 
		\hline
		LR&	1.30 & 6.27 & 1.00 & 19.11 & 1.52 & 0.53 \\ 
	
	\end{tabular}
\end{table}

Note that mixing across subpopulations is not the same as averaging the allele frequencies and assuming an
undivided subpopulation. However, the appropriate analysis based on mixing subpopulations as in \textcite{green:mortera:09} needs to be adapted to DNA mixtures and is beyond the scope of this paper.

\paragraph{Upper bound on the \LR.}
For a single source trace the weight of evidence, WoE is simply $-\log_{10} \pi_s$, where $\pi_s= P(U=s)$  is the \emph{match probability}   \ie\ the probability  that a random member of the population has the  specific DNA profile of a suspect $s$.
\begin{table}[ht!]
	\centering
	\caption{Upper bound on $\mbox{WoE}=-\log_{10} \pi_A$ for suspect $A$.}
	\label{tab:piA}
	\begin{tabular}{r|ccccc}
		Reference Population &   IT    &    MA &         PO &         ES \\
		\hline	
		Bound & 15.89 & 14.66 & 15.25&  16.14 \\
		
	\end{tabular}
\end{table}

\begin{table}[ht!]
	\centering
	\caption{Upper bound $\mbox{WoE}=-\log_{10} \pi_B$ for PoI $B$.}
	\label{tab:piB}
	\begin{tabular}{r|cccc}
		Reference Population &   IT    &    MA &         PO &         ES \\
		\hline
		Bound & 17.70	& 16.47 & 15.86&  16.30\\	
	\end{tabular}
\end{table}

We  point out that  $\mbox{WoE}=\log_{10}\LR\leq -\log_{10} \pi_s$ implying that a mixed trace can never give stronger evidence than a high-quality trace from a single source.

\tabref{piA} (\tabref{piB}) gives the upper bound $-\log_{10} \pi_A$ ($-\log_{10} \pi_B$) on the weight of evidence which  is much larger than the largest $\mbox{WoE}= \log_{10} \LR=\log_{10} 1.8=0.26$ in \tabref{ALR}  and \tabref{ALRunc} ( $\mbox{WoE}=1.73$ in \tabref{BLR} and \tabref{BLRunc}). 

The \emph{loss of evidential efficiency} $\mbox{WL}(E\cd A)$ against the suspect having genotype $A$ for evidence $E$ 
\[\mbox{WL}(E\cd A)= 
- \log_{10}\pi_A - \log_{10}\frac{\Pr(E\cd H_p)}{\Pr(E\cd H_d)}.
\]
gives the number of  bans that are lost due to the evidence being based on a mixture rather than a single source trace. For example, when using the Macedonian Romani reference population and mixture B3,  the  $\mbox{WL}(E\cd A)= 14.3$ is large. This  points out that the data in this case were far from incriminating for the suspect $A$.

\bibliographystyle{oupvar}
\bibliography{dna,sensitivity, refs}

\section*{Appendix 1}
\paragraph{Some historical facts about the Romani Population}
The Fascist regime in Italy marked  ``Gypsies'' as both  ``undesirable foreigners'' and  ``dangerous Italians'', thereby creating a dual rationale for placing them in police confinement and interning them after Italy's entrance into the war \cite{trevisan2017}. 

Under Nazi Germany, a supplementary decree to the Nuremberg Laws was issued on 26 November 1935, classifying Gypsies as ``enemies of the race-based state'', thereby placing them in the same category as the Jews. The Romani genocide or the Romani Holocaust was the effort by Nazi Germany and its World War II allies to commit genocide against Europe's Romani people. Thus, in some ways the fate of the Roma in Europe paralleled that of the Jews. 
Historians estimate that between 220,000 and 500,000 Romani were killed by the Germans and their collaborators -- 25\% to over 50\% of the slightly fewer than 1 million Roma in Europe at the time.

An excerpt of the Italian  ministerial internment order of September 11, 1940 reads:  ``...  due  to  the  fact  that  they  sometimes  commit  serious  crimes  because  of  their  innate nature  and  methods  of  organisation  and  due  to  the  possibility  that  among  them  there are elements capable of carrying out anti-national activities, it is indispensable that all Gypsies are controlled ... It is ordered that those of Italian nationality, either confirmed or  presumed,  who  are  still  in  circulation  are  to  be  rounded  up  as  quickly  as  possible and concentrated under vigorous surveillance in a suitable locality in every province ... apart from the more dangerous or suspicious elements who are to be sent to the islands ... ''.

The Italian  Ministry  of  Interior in 1940 ordered that the camps were to be established in derelict or rarely used buildings, far from strategically important centres and wherever possible in remote areas. Most of the camps were in the regions of central Italy, particularly in the central Apennine valley and the Abruzzi \cite{boursier1999}.
\newpage
\section*{Appendix 2}
 \begin{table}[ht!]
	\centering
	\label{my-label}
	\begin{tabularx}{\textwidth}{X}
		\toprule
{	\textbf{Algorithm:} Presence Index computation} \\ \midrule
		\textbf{Step 1:} \\
		\ Select a {\it contributor} and a DNA mixture  sample. \\
		\textbf{Step 2:} \\
For every marker  $m = 1,..., M$  assign the following values:\\
\begin{itemize}
	\item 0: if none of the contributor's alleles has a peak height in the mixture greater than $C$. 
	\item 0.5: if a heterozygous contributor has only one of his alleles has a  peak height in the mixture greater than $C$.
	\item 1: for a heterozygous contributor, if both his alleles have a peak height in the mixture greater than $C$; or, for a homozygous contributor, if this allele has a peak height in the mixture greater than $C$.
	
\end{itemize}
	\textbf{Step 3:} \\
	Sum the quantities obtained at each marker and divide by $M$.
	 \\ \bottomrule
	\end{tabularx}
\end{table}
\end{document}